\documentclass[reprint,superscriptaddress,amssymb,amsmath,aps,pra]{revtex4-1}

\usepackage{graphicx}
\usepackage{amssymb}
\usepackage{epstopdf}
\usepackage{upgreek}
\usepackage{dcolumn}
\usepackage{bm}
\usepackage{gensymb}
\usepackage{braket}
\usepackage{amsmath}
\usepackage{float}

\usepackage{tikz}

\expandafter\ifx\csname package@font\endcsname\relax\else
 \expandafter\expandafter
 \expandafter\usepackage
 \expandafter\expandafter
 \expandafter{\csname package@font\endcsname}%
\fi

\newcommand{\microamp}{~$\upmu$A}

\newcommand{\be}{\begin{eqnarray}}
\newcommand{\ee}{\end{eqnarray}}
\newcommand{\bfig}{\begin{figure}}
\newcommand{\efig}{\end{figure}}

\DeclareFontFamily{U}{mathb}{}
\DeclareFontShape{U}{mathb}{m}{n}{
  <-5.5> mathb5
  <5.5-6.5> mathb6
  <6.5-7.5> mathb7
  <7.5-8.5> mathb8
  <8.5-9.5> mathb9
  <9.5-11.5> mathb10
  <11.5-> mathbb12
}{}
\DeclareRobustCommand{\sqcdot}{%
  \mathbin{\text{\usefont{U}{mathb}{m}{n}\symbol{"0D}}}%
}

\usepackage[colorlinks=true,allcolors=blue]{hyperref}%

\begin{document}

\title{Optimal operation of a Josephson parametric amplifier for vacuum squeezing}
\author{M. Malnou}
\email{maxime.malnou@colorado.edu}
\affiliation{JILA, National Institute of Standards and Technology and the University of Colorado, Boulder, Colorado 80309, USA}
\affiliation{Department of Physics, University of Colorado, Boulder, Colorado 80309, USA}
\author{D. A. Palken}
\affiliation{JILA, National Institute of Standards and Technology and the University of Colorado, Boulder, Colorado 80309, USA}
\affiliation{Department of Physics, University of Colorado, Boulder, Colorado 80309, USA}
\author{Leila R. Vale}
\affiliation{National Institute of Standards and Technology, Boulder, Colorado 80305, USA}
\author{Gene C. Hilton}
\affiliation{National Institute of Standards and Technology, Boulder, Colorado 80305, USA}
\author{K.~W. Lehnert}
\affiliation{JILA, National Institute of Standards and Technology and the University of Colorado, Boulder, Colorado 80309, USA}
\affiliation{Department of Physics, University of Colorado, Boulder, Colorado 80309, USA}
\date{\today}

\date{\today}

\begin{abstract}
A Josephson parametric amplifier (JPA) can create squeezed states of microwave light, lowering the noise associated with certain quantum measurements. We experimentally study how the JPA's pump influences the phase-sensitive amplification and deamplification of a coherent tone's amplitude when that amplitude is commensurate with vacuum fluctuations. We predict and demonstrate that by operating the JPA with a pump power greater than the value that maximizes gain, the amplifier distortion is reduced and consequently squeezing is improved. Optimizing the JPA's operation in this fashion, we directly observe $3.87 \pm 0.03$ dB of vacuum squeezing. 

\end{abstract}

\maketitle

\section{Introduction}

Josephson parametric amplifiers (JPAs) have enabled many advances in quantum information measurements, such as the observation of quantum jumps \cite{vijay2011observation,lin2013single,hatridge2013quantum}, fast qubit initialization \cite{riste2012initialization}, and characterization of microwave single photon sources \cite{kindel2016generation}. When operated in a phase-sensitive mode, a JPA amplifies one quadrature of the electromagnetic field while deamplifying the other, enabling the squeezing of vacuum states below the standard quantum limit. Such squeezed states can improve qubit readout fidelity in dispersive measurements \cite{murch2013reduction,barzanjeh2014dispersive,didier2015heisenberg,didier2015fast}, help to better measure the motion of a mechanical oscillator \cite{clark2016observation}, and enhance the signal-to-noise ratio of spin-echo detection in a magnetic resonance experiment \cite{bienfait2016magnetic}. Similarly, they could accelerate haloscope-based searches \cite{sikivie1983experimental} for axionic dark matter \cite{zheng2016accelerating}, an application in which JPAs have already been successfully deployed, albeit only in a phase-insensitive mode of operation \cite{brubaker2017first,alkenany2017design}.


In principle, a JPA can squeeze one quadrature of vacuum while amplifying the other by the same amount \cite{castellanos2008amplification,mallet2011quantum,menzel2012path}. However, in practice vacuum squeezing is limited by microwave losses and distortion due to amplifier saturation \cite{boutin2017effect,liu2017josephson}. Minimizing loss through improved, low-loss elements or via fewer microwave connections remains an important challenge, and, to that end, on-chip circulators \cite{sliwa2015reconfigurable,Kerckhoff2015onchip,lecoq2017nonreciprocal,mahoney2017onchip,chapman2017widely}, flux-pumped parametric amplifiers \cite{yamamoto2008flux,zhong2013squeezing,zhou2014high,pogorzalek2017hysteretic}, and directional amplifiers \cite{sliwa2015reconfigurable,macklin2015near} are promising. We here address the other challenge: distortion.

A JPA is essentially a nonlinear $LC$ resonator whose nonlinearity comes from the Josephson junctions comprising its inductance \cite{castellanos2007widely,castellanos2008amplification}. Parametric oscillation of the circuit converts a pair of pump photons at angular frequency $\omega_p$ into a signal photon at $\omega_s$ and a complementary idler photon at $\omega_i$ in a four-wave process: $2\omega_p = \omega_s + \omega_i$. In the degenerate case, the pump is centered about the amplification band and $\omega_p \simeq \omega_s \simeq \omega_i$.

Different combinations of pump power and frequency will produce the same gain, and one typically operates at points first order-insensitive to pump power fluctuations, i.e. on the line of maximum gain (LMG) \cite{manucharyan2007microwave}. Here a JPA can achieve more than 20 dB of gain and maintain a linear relationship between input and output signal powers. Squeezing, however, is more susceptible to nonlinearities that distort the output, arising when signal photons in the resonator greatly affect the pump's amplitude. These nonlinear effects impact the squeezed quadrature at gains much lower than those where they affect the amplified quadrature. Pump stiffness can nonetheless be improved by double-pumping \cite{kamal2009signal} or flux-pumping \cite{yamamoto2008flux,zhong2013squeezing,zhou2014high,pogorzalek2017hysteretic}, at the cost of increased design and control complexity.


In this article, we demonstrate how a single-pumped JPA can be optimized for squeezing. We show that pumping it with a power greater than that which maximizes gain at a given pump frequency improves pump stiffness and minimizes signal distortion. We first experimentally study this effect on a coherent tone whose power is comparable to vacuum, and measure how it is transformed by a JPA operated at constant gain, for various pump powers and frequencies. Then, we determine an optimal operating point for which the distortion is minimized. Finally, we use a second JPA as a phase-sensitive preamplifier to measure a vacuum squeezed state prepared by the first JPA and show that squeezing follows the same dependence on operating point. With optimized pump power and frequency, and while keeping microwave losses low, we generate, transport, and readout a state $3.87 \pm 0.03$ dB below vacuum. This could more than double the speed at which the search for axionic dark-matter is carried out \cite{zheng2016accelerating}.

\section{Signal transformation in phase space}
\label{sec:phase_space}


In the JPAs we study, the inductance $L$ is implemented with a series array of $N_s$ balanced dc superconducting quantum interference devices (SQUIDs). When shunted by a capacitor $C$, it forms a nonlinear resonant circuit whose frequency is tunable with magnetic flux applied through the SQUID loops. Such a tunable Kerr circuit (TKC) is described to leading nonlinear order by the Hamiltonian

\begin{equation} \label{H Kerr}
\hat{H} = \hbar\omega_0 \hat{n} +  \frac{\hbar K}{2} \hat{n}^2,
\end{equation}
which gives rise to a Duffing oscillator equation of motion \cite{manucharyan2007microwave}. Here $\hat{n} = \hat{a}^\dagger \hat{a}$ gives the number of photons in the cavity, $K = \frac{-\hbar\omega_0^2}{16N_s\phi_0I_c}$ is the Kerr nonlinearity, $I_c$ is the critical current of each junction, $\omega_0 = 1/\sqrt{LC}$ is the TKC's bare resonance angular frequency, and $\phi_0$ is the reduced magnetic flux quantum.

\begin{figure}[!h]
  \centering
    \includegraphics[scale=0.34]{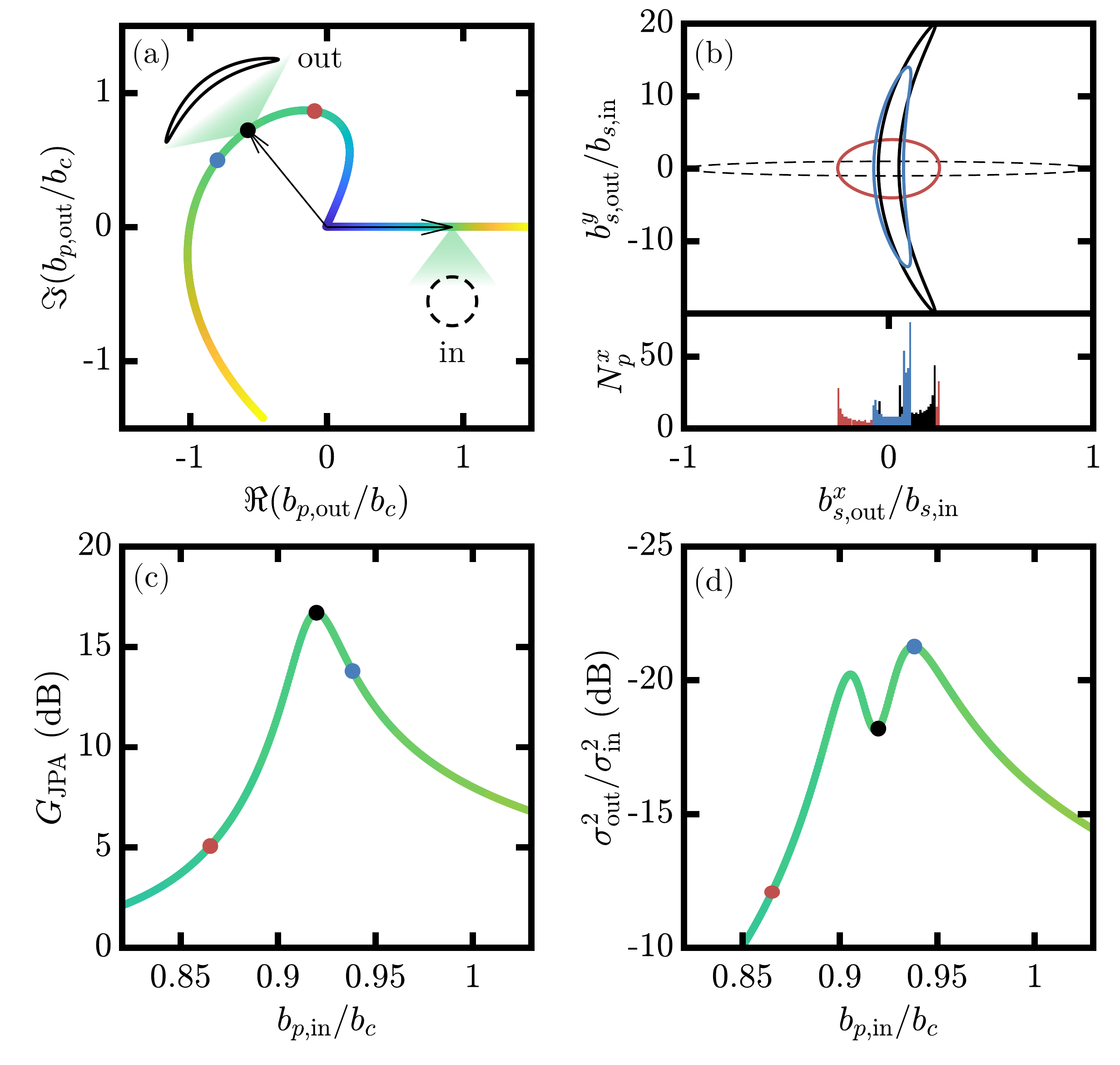}
      \caption{Theory of distortion in a JPA. Calculations for $\gamma = 2\pi\times54.5$ MHz, $\Delta/\gamma = 1.54$, $K/\gamma= -8.3 \times10^{-4}$, and a quality factor $Q=\omega_0/2\gamma = 65$, typical values for our array-based JPAs. (a) A family of input pump phasors of different amplitudes describes the horizontal, colored line. These inputs map onto output phasors distributed along the correspondingly colored curve in the complex plane. They are normalized by the critical amplitude $b_c = \sqrt{\frac{4\gamma^2}{3\lvert K \rvert\sqrt{3}}}$. Two black arrows indicate the pair of input and output phasors at maximum gain. At this operating point, a zoom-in shows how an input signal phasor $b_{s,\text{in}}e^{i\theta}$ maps onto a banana at the output. Two other operating points (red and blue) are indicated. (b) The input phasor, simulated with 360 different $\theta$ angles (dashed circle, deformed due to the unequal axis scaling), is processed by the JPA for three different pump powers corresponding to the operating points from (a). The lower panel shows the distribution of points $N^x_p$ of these phasors, when projected onto the deamplified x-axis. The JPA direct gain (c) and the deamplified quadrature variance (d) are plotted as a function of pump amplitude.}
\label{fig:theory}
\end{figure}

When coupled to an external port at field-decay rate $\gamma$, a classical limit of input-output theory \cite{gardiner1985input,yurke2006performance} determines the complex field amplitude, or phasor, at the output $b_\text{out}$ from the incoming field amplitude $b_\text{in}$ as 
\begin{equation} \label{in_out}
b_\text{out} = b_\text{in}\left(1 - \frac{2\gamma}{i(\Delta + Kn) + \gamma}\right).
\end{equation}
Here, $n = \braket{\hat{n}}$ is the average photon number and $\Delta = \omega_0 - \omega_p$ is the detuning between the bare resonant angular frequency and the input phasor's angular frequency. A JPA is often approximated as amplifying an infinitesimally weak signal $b_{s,\text{in}}e^{i\theta}$ in the presence of a large pump tone $b_{p,\text{in}}$ (taken as phase reference), such that $b_\text{in} = b_{p,\text{in}} + b_{s,\text{in}}e^{i\theta}$. However, Eq.\,\eqref{in_out} remains valid without the stiff pump approximation, providing a mapping between input and output within the bandwidth of the lossless JPA. This transformation being unitary below bifurcation, the gain derives from the sharp dependence of the output phase on the input amplitude. The photon number $n$ is determined by the cubic equation \cite{yurke2006performance}
\begin{equation} \label{n cubic}
n^3 + \frac{2\Delta}{K}n^2 + \frac{\Delta^2 + \gamma^2}{K^2}n = \frac{2\gamma}{K^2} |b_\text{in}|^2.
\end{equation}
Figure \ref{fig:theory}a shows how a family of real input pump tones with varying amplitude $b_{p,\text{in}}$ maps onto complex outputs $b_{p,\text{out}}$. They are calculated from Eqs.\,\eqref{in_out} and \eqref{n cubic}, with $\Delta$ constant. A particular input phasor and its corresponding output are indicated with black arrows, illustrating the JPA's behavior at the greatest phase sensitivity, i.e. at maximum gain.

When adding a signal $b_{s,\text{in}}e^{i\theta}$, the total input phasor $b_\text{in}$ undergoes a $\theta$-dependent transformation. One quadrature of the output signal $b_{s,\text{out}}$ is amplified while the orthogonal one is deamplified. At low JPA gain or infinitesimal input signal, $b_{s,\text{out}}$ exhibits an elliptic shape. However, as the input signal or gain increases, it distorts into a ``banana" indicating that the pump is no longer stiff. Figure \ref{fig:theory}b represents such distortions at three different JPA gains obtained by varying the pump's amplitude, for a probe tone whose power is equivalent to half a photon over the JPA's bandwidth. They are rotated to share their amplified and deamplified quadratures, respectively $b^y_{s,\text{out}}$ and $b^x_{s,\text{out}}$. At the point of maximum gain, $b_{s,\text{out}}$ actually has a larger projection on the deamplified axis $x$ than at intermediate gain, where the projection onto this axis is minimal. When a signal is detuned from the pump, it creates output rotating phasors such as these, and the direct, or signal power, gain is expressed as \cite{yurke1989observation} 
\begin{equation} \label{direct_gain}
G = \frac{\braket{G_\theta} + 1}{2},
\end{equation}
with $\braket{G_\theta}$ the average ratio of output to input rotating phasor power. Figure \ref{fig:theory}c shows the JPA direct gain as a function of pump amplitude $b_{p,\text{in}}$.

A vacuum state, considered semi-classically as a statistical ensemble of phasors, should undergo similar distortion when squeezed by a JPA. This effect degrades the maximal achievable squeezing, because it increases the noise variance along the deamplified quadrature axis. To make a qualitative estimate of the attainable degree of squeezing, we first project the distributions of input and output points onto the deamplified quadrature axis (Fig.\,\ref{fig:theory}b), and then compare the variances, ($\sigma^2_\text{in}$) and ($\sigma^2_\text{out}$), of these projections. Figure \ref{fig:theory}d shows the ratio ${\sigma^2_\text{out}}/{\sigma^2_\text{in}}$, as a function of pump power, for a given pump frequency. Deamplification is maximized when this ratio is minimized. Strikingly, the pump amplitude that produces maximum gain fails to yield maximum deamplification due to phasor-distortion of the vacuum-sized input. Instead, deamplification reaches a global maximum at an amplitude greater than that at maximum gain, i.e. above the LMG, where phasor distortion is better mitigated. 


This benefit can be understood qualitatively from the dependence of the effective frequency $\omega_\text{eff}=\omega_0 + n K / 2$ on incident amplitude $b_\text{in}$. The signal $b_{s,\text{in}}$ periodically increases $b_\text{in}$, changing the energy in the resonator for two reasons: first the incident power $\lvert b_\text{in} \rvert^2$ is larger, and second, $\omega_\text{eff}$ decreases, in turn altering the effective detuning $\Delta_\text{eff} = \omega_\text{eff} - \omega_p$. Below the LMG these effects add, but above they partially cancel; thus the gain is less dependent on signal amplitude above the LMG. We  demonstrate  this  behavior in Sec.\,\ref{sec:deamp}, enabling the optimization of vacuum squeezing discussed in Sec.\,\ref{sec:squeezing}.




\section{Deamplification of a coherent tone}
\label{sec:deamp}

\begin{figure}[!h]
  \centering
    \includegraphics[scale=0.53]{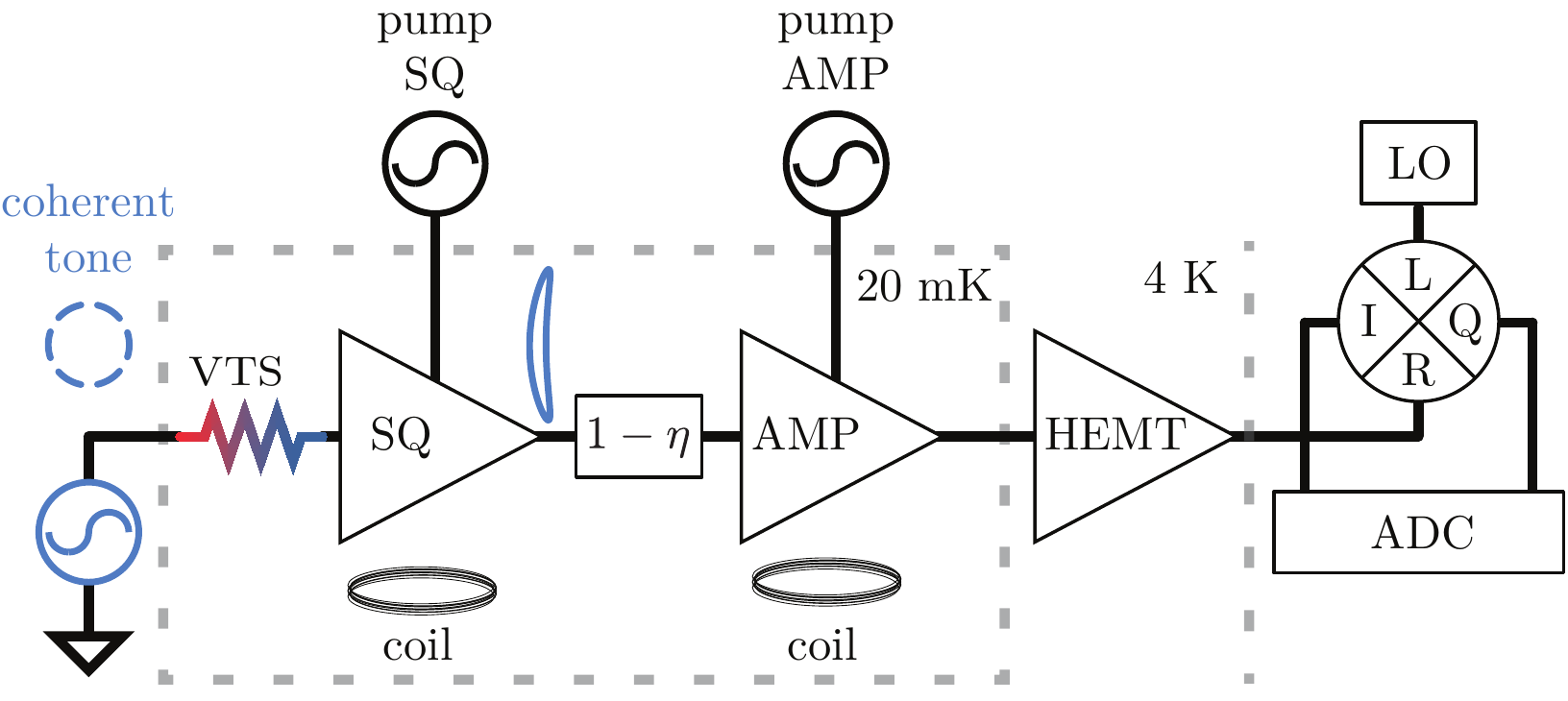}
      \caption{Experimental setup for coherent tone distortion and vacuum squeezing. The SQ and AMP are mounted in series. When studying distortion, the AMP is detuned and not pumped, while a microwave generator sends a known coherent tone at the chain's input (dashed circle), then processed by the SQ. Microwave losses between SQ and AMP are parametrized by $\eta$. After subsequent amplification by a cryogenic high electron mobility transistor (HEMT) the tone is mapped to low frequency using an IQ mixer and a local oscillator (LO). An analog-to-digital converter (ADC) records the output signal's I and Q quadratures.}
\label{fig:schematic}
\end{figure}

In a first experiment, we investigate the distortion of a vacuum-sized coherent tone when varying the JPA operating point, by changing the pump power and frequency. A single cryogenic experimental setup is used for studying distortion and vacuum squeezing, discussed in Sec.\,\ref{sec:squeezing}. It comprises two identical JPAs (appendix \ref{app:setup}) connected to each other, as shown in Fig.\,\ref{fig:schematic}. The first we label the squeezer (SQ), and the second the amplifier (AMP). Each can be independently tuned via its own superconducting coil, and a variable temperature stage (VTS) is used to inject known thermal noise into the amplifier chain and measure the SQ added noise (appendix \ref{app:noise}). We use microwave generators to pump the two JPAs and send coherent tones to the input of the chain. An IQ mixer demodulates the output quadratures with respect to a local oscillator (LO). In this first experiment, we don not use the AMP; by tuning it more than 1 GHz below the SQ band, it acts as a passive element of unit gain.

\begin{figure}[!h]
  \centering
    \includegraphics[scale=0.34]{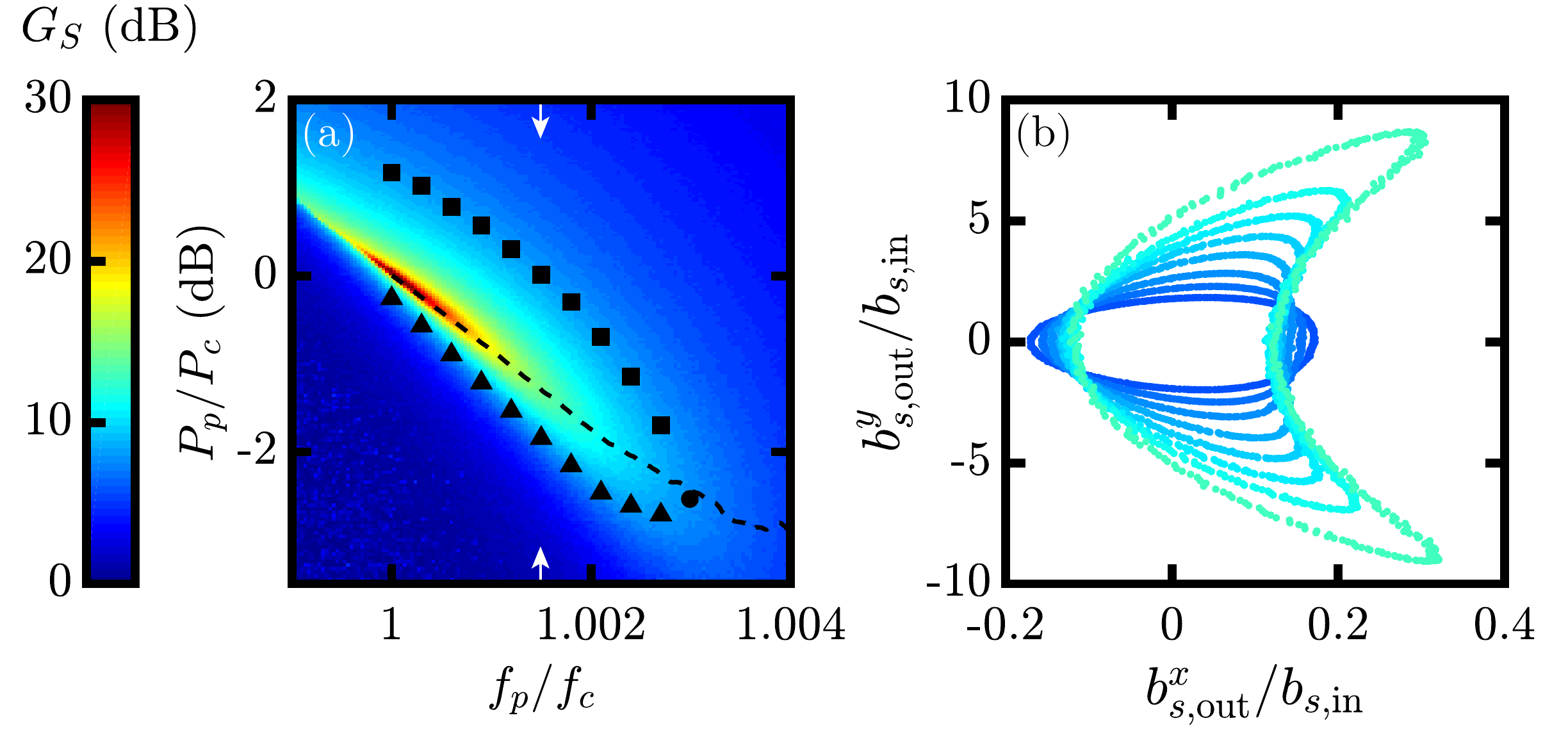}
      \caption{Dependence of coherent tone distortion upon amplifier operating point. (a) The SQ gain map, with critical frequency $f_c = 7.0032$ GHz and input critical power $P_c = - 93.1$ dBm. (b) Output tone in the IQ-plane (input is a circle of unit radius, not shown). The SQ direct gain is varied between 6 (blue) and 13 dB (green), and operated above the LMG at $f_p/f_c = 1.0015$, a frequency indicated by arrows in (a).}
\label{fig:deamplification_N1}
\end{figure}

In order to compare distortion and SQ direct gain $G_S$ at particular operating points, we find $G_S$ (measured with a vector network analyzer, see appendix \ref{app:setup}) as a function of pump power $P_p$ and frequency $f_p$. We identify a critical point at which the gain diverges, and describe the pump power and frequency as fractions of these critical values (Fig.\,\ref{fig:deamplification_N1}a). The dashed black line shows the LMG. We compare deamplification when operating the SQ at points with equal gains below, above, and on this line.

To visualize phasor distortion, we now input a vacuum-sized coherent tone $b_{s,\text{in}}e^{i\theta}$ whose phase $\theta$ slowly rotates in the SQ pump's frame (appendix \ref{app:setup}). With LO and SQ pumps at the same frequency, the tone describes traces at the output in the IQ-plane such as those in Fig.\,\ref{fig:deamplification_N1}b. They are obtained with the SQ operated above the LMG. Thus, decreasing the pump power, $G_S$ increases from 6 to 13 dB, as the output transitions from ellipse to banana.

\begin{figure}[!h]
  \centering
    \includegraphics[scale=0.34]{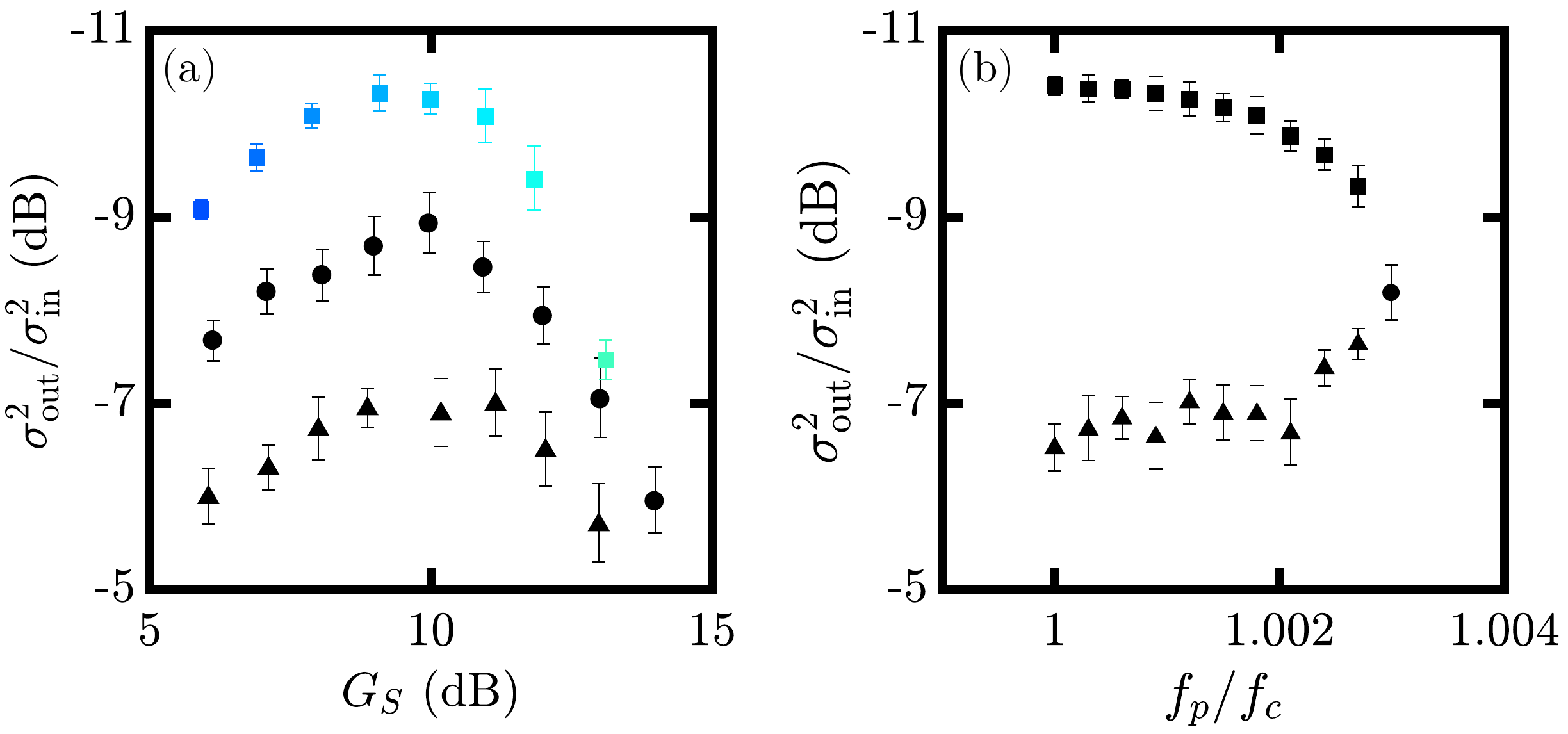}
      \caption{Dependence of deamplification upon amplifier operating point. In (a) deamplification is shown as a function of SQ direct gain when the SQ is operated below ($\blacktriangle$) and above ($\sqcdot$) the LMG (at $f_p/f_c = 1.0015$), as well as on ($\bullet$) the LMG (where the pump frequency has been varied). In (b) it is shown for the black points in Fig.\,\ref{fig:deamplification_N1}a, where $G_S = 8$ dB.}
\label{fig:deamplification_N2}
\end{figure}

Upon repeating the measurement with the SQ off, we characterize deamplification as a function of the SQ operating point by forming the ratio ${\sigma^2_\text{out}}/{\sigma^2_\text{in}}$, as seen in Sec.\,\ref{sec:phase_space}. Figure \ref{fig:deamplification_N2}a shows these ratios, calculated from the points in Fig.\,\ref{fig:deamplification_N1}b. Compared to ratios obtained for operating points below and on the LMG for similar sets of JPA gains, we see that deamplification increases for increasing SQ direct gains less than $G_S = 9$ dB, as the ellipse stretches. Above $9$ dB, it decreases due to phasor distortion.

Finally, we study deamplification along a contour of constant direct gain, encircling the critical point, to look for an optimal operating point at which distortion is minimized. Figure \ref{fig:deamplification_N2}b characterizes the tone's deamplification along the $G_S=8$ dB contour, at operating points away from bifurcation (i.e. for $f_p > f_c$), indicated in black in Fig.\,\ref{fig:deamplification_N1}a. Below the LMG ${\sigma^2_\text{out}}/{\sigma^2_\text{in}}$ is about $-7$ dB, whereas above it decreases to about $-10$ dB, as a result of reduced distortion. Thus, operating the SQ above the LMG and with a gain of about 9 dB maximizes deamplification. We leverage these findings in Sec.\,\ref{sec:squeezing} to optimally bias the SQ.

\section{Optimizing vacuum squeezing} 
\label{sec:squeezing}

Most applications of microwave squeezed light require the ability to both generate and efficiently detect a squeezed state. We demonstrate these dual capabilities with the AMP, cascaded after the SQ. Upon transport from SQ to AMP, the state suffers from microwave loss $\eta$ (Fig.\,\ref{fig:schematic}), which replaces part of the state with unsqueezed vacuum \cite{leonhardt1995measuring}, diminishing the measurable degree of squeezing. We reduced these losses to $\eta=1.2\pm0.2$ dB in our setup (appendix \ref{app:lines}) using superconducting cables and a narrowband triple-junction circulator with insertion loss $<0.5$ dB.

To measure vacuum squeezing, we pump the AMP to get more than 20 dB of gain, in order to overcome the amplification chain's added noise. Then, the SQ pump frequency is detuned from the AMP pump by 20 kHz. Thus, the squeezed quadrature rotates in the AMP frame, and is periodically amplified. Because SQ and AMP operate at nearby frequencies, both pumps need to be canceled. This avoids AMP saturation and preserves squeezing of a pure vacuum state. We perform these cancellations by combining $\pi$-phase-shifted pump tones at the output of each JPA (appendix \ref{app:setup}). Finally, the VTS is operated at a temperature $T_\text{VTS} = 55$ mK, i.e. $k_B T_\text{VTS} \ll \hbar\omega_S$ such that the SQ input sees only quantum noise.

\begin{figure}[!h]
  \centering
    \includegraphics[scale=0.34]{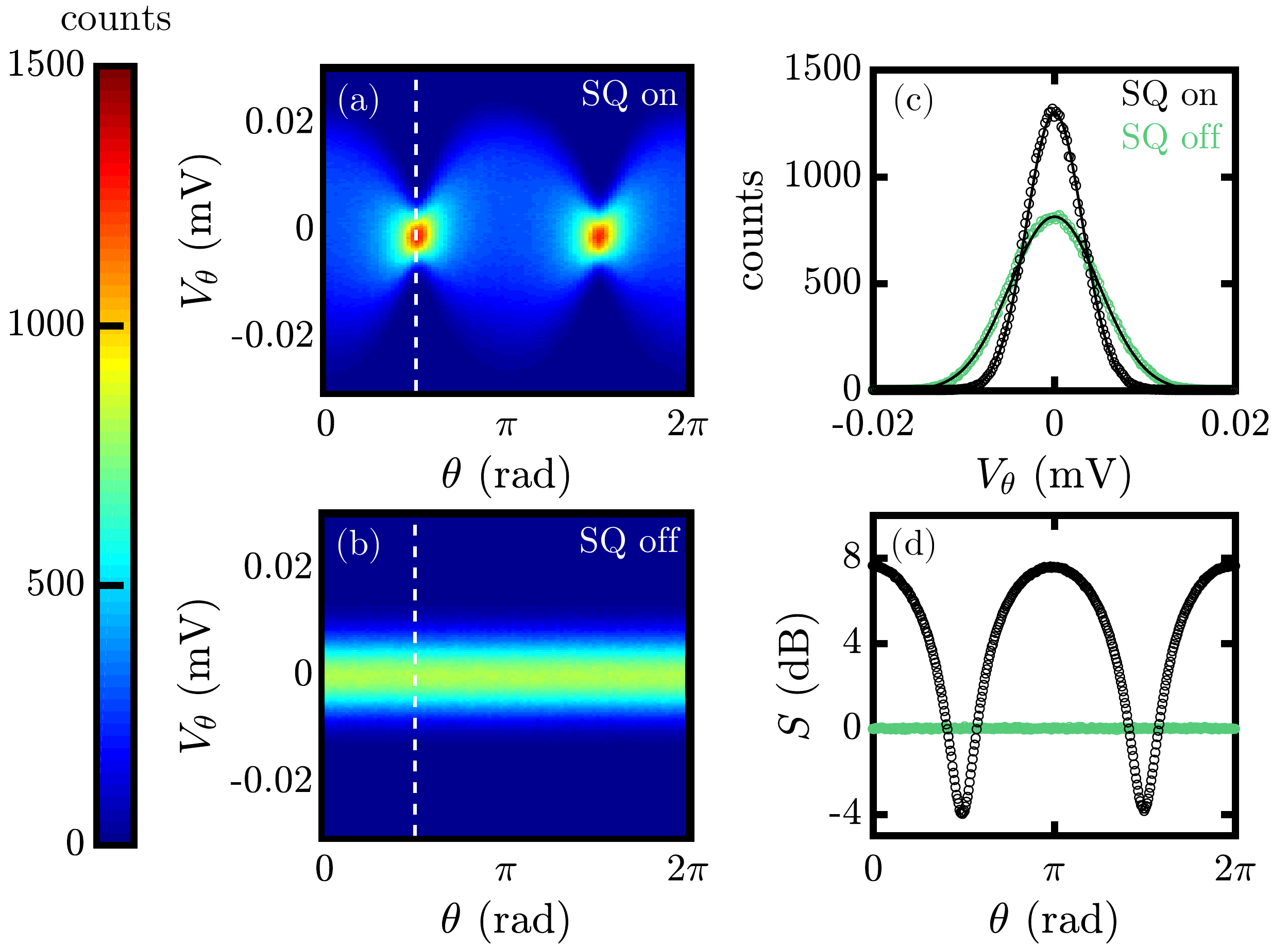}
      \caption{Vacuum squeezing. The SQ is operated with direct gain $G_S = 8$ dB, obtained with a pump frequency $f_p/f_c = 1.0015$ and pump power $P_p/P_c = 0.5045$ dB. (a) Histogram of output voltage fluctuations $V_\theta$ along the AMP amplified quadrature with SQ on as a function of the phase difference $\theta$ between SQ and AMP pumps. (b) Histogram with SQ off. (c) Profile of the fluctuations and Gaussian fits (solid lines) along the dashed lines drawn in panels (a) and (b), at $\theta=\pi/2$. (d) Squeezing $S=\sigma^2_\text{on}/\sigma^2_\text{off}$ as a function of $\theta$.}
\label{fig:squeezing}
\end{figure}

We operate the SQ above the LMG, with a pump frequency $f_p/f_c = 1.0015$, and with direct gain $G_S=8$ dB, in order to maximize squeezing, as expected from Sec.\,\ref{sec:deamp}. Fig.\,\ref{fig:squeezing}a shows a histogram of the voltage fluctuations $V_\theta$ retreived along the AMP amplified quadrature as a function of  $\theta$, the phase difference between the SQ and AMP pumps. Compared to panel (b), where the SQ is off, vacuum has been squeezed for $\theta=\pi/2$ and $3\pi/2$. Figure \ref{fig:squeezing}c presents a cut in the histograms at $\theta=\pi/2$, demonstrating a clear reduction in the Gaussian's standard deviation, and Fig.\,\ref{fig:squeezing}d displays the squeezing $S = \sigma^2_\text{on}/\sigma^2_\text{off}$ as a function of $\theta$, with a minimum of $-3.87\pm 0.03$ dB.

\begin{figure}[!h]
  \centering
    \includegraphics[scale=0.34]{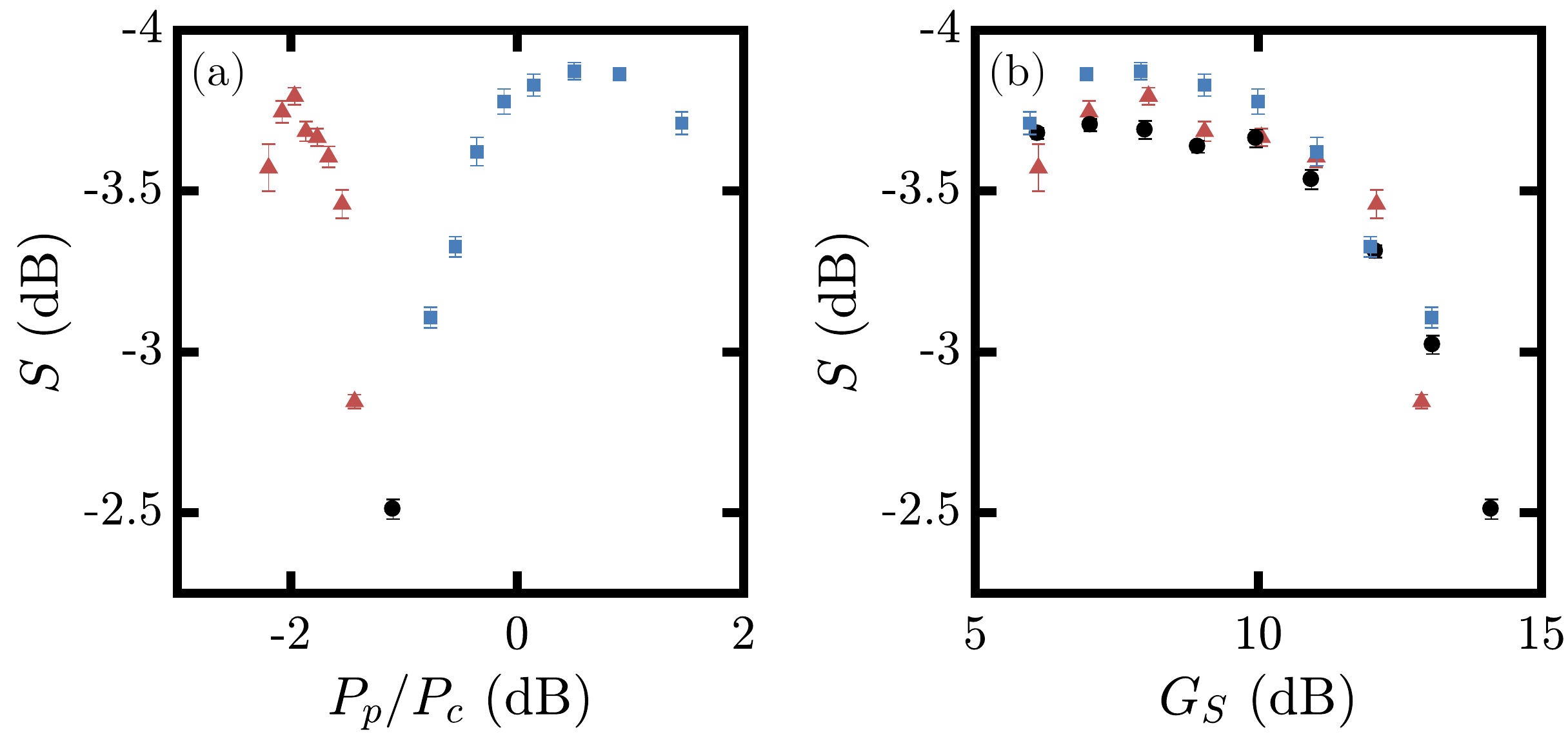}
      \caption{Dependence of vacuum squeezing on operating point. (a) The points in red, black, and blue are, respectively, below, on, and above the LMG. Squeezing is plotted as a function of pump power, for $f_p/f_c = 1.0015$. In (b) it is plotted as a function of SQ gain, and the red and blue points are those used in (a). For the black points, $f_p$ has been varied to obtain gains between 6 and 14 dB.}
\label{fig:squeezing_JPAgain}
\end{figure}

Moving the SQ operating point, we verified that squeezing decreases as we approach the LMG, due to increasing gain. Adjusting the pump's power while keeping the frequency constant, we can move the operating point from below to above the LMG, describing a line cut in the gain map, indicated by arrows in Fig.\,\ref{fig:deamplification_N1}a. Figure \ref{fig:squeezing_JPAgain}a shows the squeezing along this cut, in qualitative agreement with the theoretical and experimental results on coherent tones seen in Secs.\,\ref{sec:phase_space} and \ref{sec:deamp}. Furthermore, we obtain a slightly better squeezing above the LMG than below, as expected. This feature is displayed in Fig.\,\ref{fig:squeezing_JPAgain}b, where squeezing is reported as a function of SQ gain. As losses are further reduced, we anticipate further improvement in performance, as observed in Sec.\,\ref{sec:deamp} for the deamplification of a coherent tone.


\section{Conclusion} 

Pump stiffness is crucial when generating a microwave squeezed state with a JPA, because any deviation from the stiff-pump regime strongly affects the squeezed quadrature. We have shown that a lack of stiffness distorts output signals, thereby limiting the maximum degree of squeezing. However, we have also presented a partial solution to this problem by means of choosing the operating point to lie above the LMG, the intuitive choice. Experiments conducted on a coherent tone transformed by a JPA confirm the efficacy of this approach, with less phasor distortion above the LMG. When squeezing vacuum, we observed the same trend. Maintaining low microwave loss, we thus generated and delivered a $3.87\pm0.03$ dB vacuum squeezed state from one JPA to another.


\section*{Acknowledgements} 

This work was supported by the National Science Foundation, under grants PHY-1607223 and PHY-1734006, and by the Heising-Simons Foundation under grant 2014-183.

\appendix

\section{Experimental setup}
\label{app:setup}

\begin{figure}[!h]
  \centering
    \includegraphics[scale=0.8]{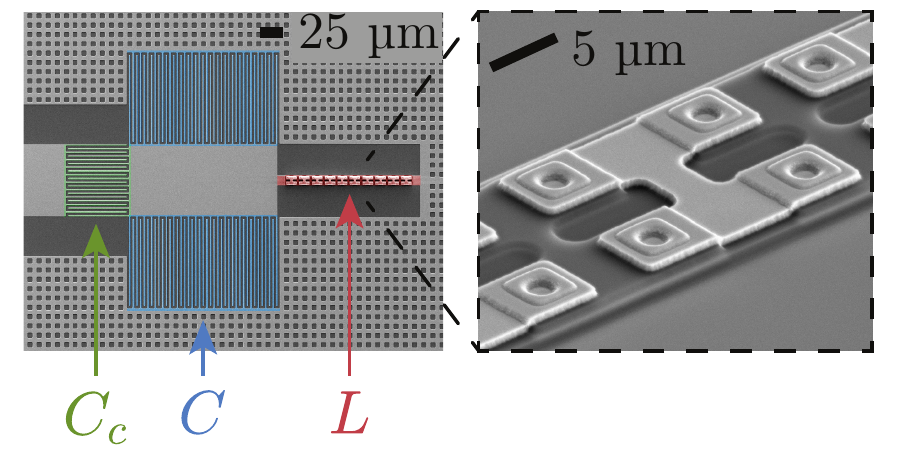}
      \caption{Scanning electron microscope images of a TKC. The chip is fabricated in a niobium-aluminum-niobium trilayer process. The waffling pattern covering the niobium ground plane pins magnetic flux trapped in the superconductor in place. A close-up view of part of the SQUID array shows four Josephson junctions, located at the ends of the bright H shape. With the base niobium electrode (dark gray), they form two SQUID loops.}
\label{fig:TKC}
\end{figure}

Our realization of a JPA comprises a TKC (Fig.\,\ref{fig:TKC}), circulator, and directional coupler. Because the TKC amplifies in reflection, we use a circulator to separate incident and reflected tones, creating input and output ports. It is pumped through the weakly coupled port of a directional coupler \cite{mallet2011quantum}. The TKC consists of a series array of $N_s = 20$ SQUIDs with junctions of critical current $I_c = 7\text{\,\microamp}$, in parallel with a $C=550$ fF interdigitated capacitor. This lumped element resonator connects through a coupling capacitance $C_c=70$ fF to a coplanar waveguide. The TKC has bare resonant frequency $f_0 = 7.3$ GHz and quality factor $Q = 65$. 

\begin{figure*}
  \centering
    \includegraphics[scale=0.6]{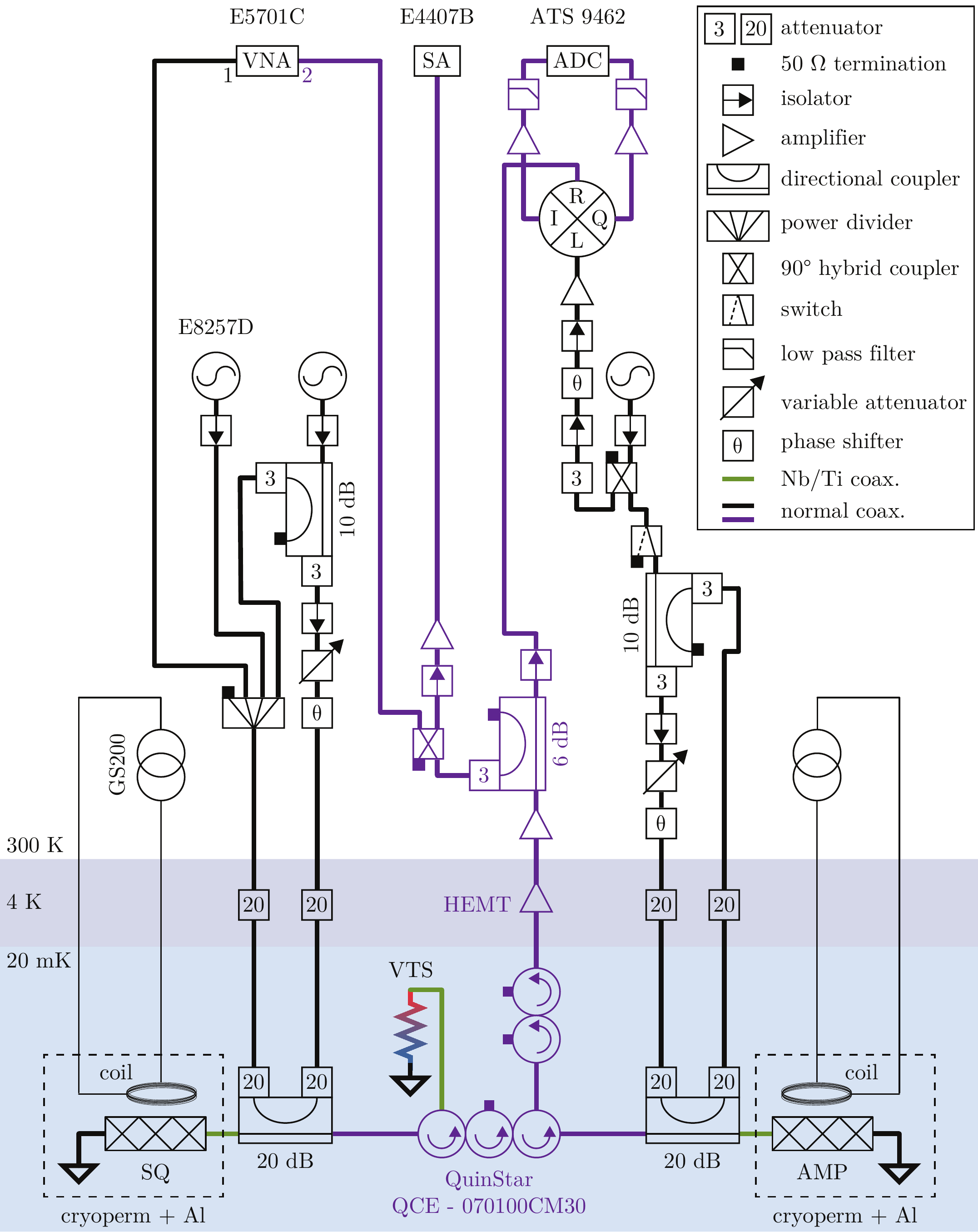}
      \caption{Full experimental setup used for both coherent tone deamplification and squeezing. The input and output coaxial cables are colored in black and purple, respectively. The green lines are superconducting.}
\label{fig:full_setup}
\end{figure*}

The full experimental setup is represented in Fig.\,\ref{fig:full_setup}. SQ and AMP TKCs are individually magnetically shielded. They can be tuned via coils biased by current sources. The generator which provides the AMP pump and cancellation tones also sources the LO. A switch may be used to turn off the AMP pump without interrupting the LO. A third generator (far left in Fig.\,\ref{fig:full_setup}) is used to create the probe tone. Output signals are demodulated with an IQ mixer, then digitized. A vector network analyzer (VNA) is used to determine the direct gains and a spectrum analyzer (SA) monitors pump cancellation. Each cancel line consists of a variable attenuator and a phase shifter. Several isolators and 3 dB attenuators are placed in the setup to minimize harmful reflections. 

At base temperature, a triple-junction circulator links the SQ and AMP TKCs. It provides $>50$ dB isolation between SQ and AMP, and has low insertion loss ($<0.5$ dB). A VTS, weakly coupled to the base temperature plate and composed of a $50\ \ohm$ load, a heater, and a thermometer, generates the chain's input noise and allows for measurement of SQ added noise. Superconducting Nb/Ti lines are placed between TKCs and DCs to further reduce microwave loss, and the DCs are directly connected to the circulator with swept elbows.

To measure the distortion of a coherent tone (Sec.\,\ref{sec:deamp}), we adjust its power to $P_\text{PR}=\frac{1}{2}\hbar\omega B$ in order to keep it comparable to vacuum, with $B$ the SQ bandwidth. With the LO generator at the same frequency as the SQ pump, we detune the tone by $f_\text{PR} - f_p = 20$ kHz, so that its phasor slowly rotates in the LO's frame, enabling the measurement of 360 independent phasors within the 15 MHz IF bandwidth of our readout chain. Due to the feeble input signal, we heavily average the two output quadratures, which then describe traces in the IQ-plane such as those in Fig.\,\ref{fig:deamplification_N1}b.

\section{Line calibration}
\label{app:lines}

Calibrating the lines relies on measuring the overall input attenuation $A^I$ from the probe tone's microwave generator to the SQ input, and also the overall amplifications $G_A^O$ and $G_S^O$ from the TKC outputs to the analog-to-digital converter (ADC). Knowing $A^I$ allows for calibration of the probe power incident on the SQ TKC, and the ratio $\eta = G_A^O/G_S^O$ provides, \textit{ceteris paribus}, an estimate of the loss experienced by the squeezed state when travelling between SQ and AMP. 

Operating one JPA at a time (without pump cancellation), $G_A^O$ and $G_S^O$ are calculated by integrating the power spectral density of vacuum fluctuations recorded by the ADC over a window $W=500$ kHz within the JPA's bandwidth. To avoid dc offsets, we perform a heterodyne measurement, with LO and JPA pump detuned by 5 MHz, and $W$ ending 400 kHz below the pump. Thus, when measuring $G_A^O$, the generator used for pumping is the one connected to the SQ, itself detuned. The SQ TKC then simply acts as a mirror, and the pump tone reaches the AMP after reflection. Given a quantum at the SQ or AMP input, the integrated power for both cases is
\begin{equation} \label{G^O}
P_{A/S}^O = \hbar\omega W G_{A/S}^O G_{A/S},
\end{equation}
with $G_{A/S}$ the AMP or SQ gain, measured with the VNA. We operated the JPAs at various gains between 20 and 30 dB, ensuring that the output spectral density linearly tracked the JPA gain, and obtained
\begin{eqnarray*}
    \text{AMP to ADC\hspace{1cm}} & G_A^O=76.5\pm0.1\text{ dB} \\
    \text{SQ to ADC\hspace{1cm}} & G_S^O=75.3\pm0.1\text{ dB}.    
\end{eqnarray*}
Therefore, we estimate that $\eta=1.2\pm0.2$ dB. In principle this value should enable us to observe about 5.6 dB of squeezing, a bit more than what we measured. But in practice, we are limited by a combination of squeezed state distortion and the contribution from the HEMT added noise, which becomes harder to overwhelm when squeezing. 



We estimated $A^I$ with a similar protocol. Turning off both JPAs, a probe tone with input power $P_P^I$ travelling in the lines creates an output power $P_P^O$ on the ADC as
\begin{equation} \label{G^I}
P_P^O = P_P^I G_S^O A^I.
\end{equation}
We thus estimated $A^I = - 81.4\pm0.2$ dB.

\section{SQ added noise}
\label{app:noise}
Any resistor at temperature $T$ generates noise of known variance, whose spectral density in units of quanta at angular frequency $\omega$ is
\begin{equation} \label{N_in}
S_\text{in} = \frac{1}{2} + \frac{1}{e^{\hbar\omega/k_BT} - 1}.
\end{equation}
Our VTS therefore allows us to generate a known thermal state at the chain's input, which is then amplified by the SQ, giving at its output:
\begin{equation} \label{NS_out}
S_{S,\text{out}} = G_S(S_\text{in} + N_S),
\end{equation}
where $G_S$ is the SQ gain and $N_S$ its added noise (in this experiment we only use the SQ; the AMP is detuned). Since we are operating the SQ in a phase-sensitive mode, $N_S\geq0$.

In practice there are inefficiencies in the thermal state's transportation from VTS to SQ. We can model these with a simple beamsplitter picture, such that a fraction $\lambda$ of $S_\text{in}$ is replaced with noise $S_f$ at the fridge's temperature. Considering the system added noise as $N_\text{add} = N_S + N_H / G_S$, where $N_H$ is the HEMT added noise, we have 
\begin{eqnarray} \label{S_out}
S_\text{out} &=& G^O_S G_S(\lambda S_{\text{in}} + (1-\lambda) S_f  + N_\text{add}) \\
&=& G^O_S G_S S_\text{out}^\text{in}
\nonumber 
\end{eqnarray}
at the output of the chain, where $G^O_S$ is the chain's gain, and $S_\text{out}^\text{in}$ is the output noise spectral density referred to the SQ input. Varying $S_\text{in}$ and $S_f$, we thus can deduce $N_\text{add}$ and $\lambda$.
\begin{figure}[!h]
  \centering
    \includegraphics[scale=0.55]{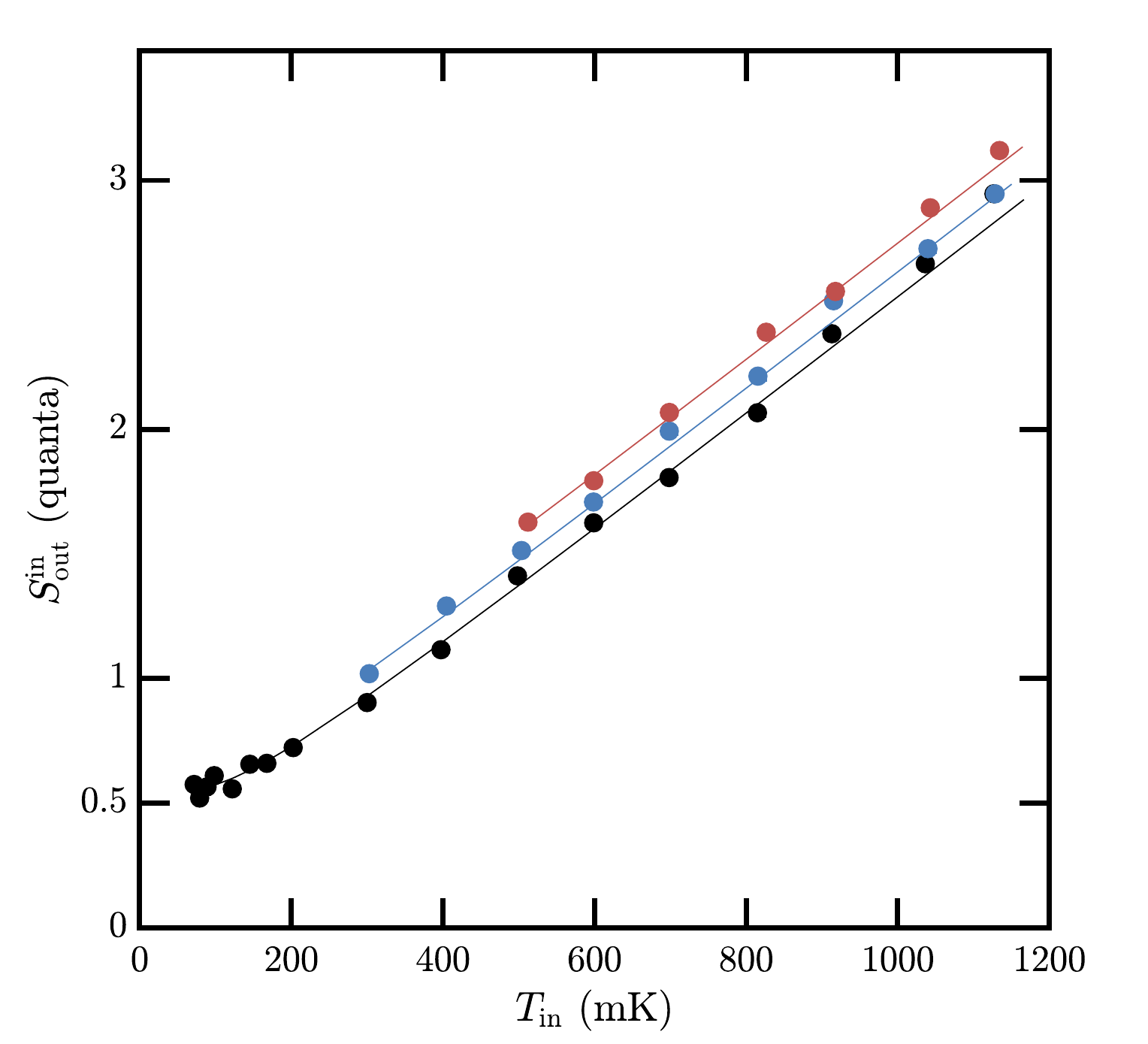}
      \caption{Noise spectral density as a function of VTS temperature. The noise is referred to the SQ input. Points in black, blue, and red are for fridge's temperatures $T_f = 50$, 300, and 500 mK, respectively. The statistical error is smaller than the point size. The solid lines are obtained from fitting the parameters $N_\text{add}$, $\lambda$ and the overall gain of the chain $G^O_S G_S$.}
\label{fig:noise_cal}
\end{figure}

With a homodyne configuration (LO and SQ pump at the same frequency), we integrate the output noise spectral density $S_\text{out}$ over the SQ bandwidth $B$ for various VTS temperatures. The fridge's base temperature $T_f$ is fixed and $G_S\geq25$ dB is tracked by amplifying and measuring a small pilot tone. Figure \ref{fig:noise_cal} shows $S_\text{out}^\text{in}$ as a function of the VTS temperature for three different fridge temperatures. A fit gives $N_\text{add} = 0.045\pm0.001$ and $\lambda = 0.79 \pm 0.01$, where the quoted uncertainties correspond to statistical errors in the measurements.

\vspace{0.1in}


%

\end{document}